\documentstyle[aps,prb,epsf]{revtex}

\newcommand{\Tk}{{\em $T_{K}$}}
\newcommand{\Tv}{{\em $T_{v}$}}

\newcommand{\Ef}{{\em $E_{f}$}}
\newcommand{\epsf}{{\em $\epsilon_{f}$}}
\newcommand{\nft}{{\em $n_{f}(T)$}}

\newcommand{\wn}{{$cm^{-1}$}}

\newcommand{\sigo}{{\em $\sigma_{1} (\omega)$}}

\newcommand{\w}{{\em $\omega$}}

\newcommand{\ybin}{YbInCu$_{4}$}

\newcommand{\ybb}{YbB$_{12}$}

\begin{document}
\twocolumn[\hsize\textwidth\columnwidth\hsize\csname@twocolumnfalse%
\endcsname
\bibliographystyle{unsrt}
\title{Optical study of the electronic phase transition of strongly correlated \ybin\ }
\author{Sean R. Garner, Jason N. Hancock, Yvonne W. Rodriguez, and Zack Schlesinger}
\address{Department of Physics, University of California\\
Santa Cruz, California 95064}
\author{Benno Bucher}
\address{Department of Physics, ITR\\
Oberseestrasse 10;  8640 Rapperswil   
Switzerland}
\author{Zach Fisk}
\address{Department of Physics and NHMFL, Florida State University\\
Talahassee, Florida 32310}
\author{John L. Sarrao}
\address{Los Alamos National Laboratory, Los Alamos\\
New Mexico, 84545}
\date{January 24, 2000}
\maketitle
\begin{abstract}
Infrared, visible and near-UV reflectivity measurements are used to obtain 
conductivity as a function of temperature and frequency in \ybin, 
which exhibits an isostructural phase-transition into a mixed-valent phase below $T_v \simeq 42 K.$
In addition to a gradual loss of spectral weight with decreasing temperature extending up to 1.5 eV,
a sharp resonance appears at 0.25 eV in the mixed-valent phase. This feature can
be described in terms of excitations into the Kondo (Abrikosov-Suhl) resonance, 
and, like the sudden reduction of resistivity, provides a direct reflection of the onset of 
coherence in this strongly correlated electron system. 
\end{abstract}
]

Phenomena in the field of strongly correlated electron systems can often be described
in terms of either a Mott-Hubbard model or a periodic Anderson model\cite{hewson} (PAM).
\ybin\ is difficult to classify in this manner.  Its low temperature properties appear to
be consistent with a PAM approach--it is mixed-valent with an enhanced carrier mass and a 
correspondingly enhanced electronic specific heat and Pauli spin susceptibility.
However, it exhibits a phase transition at 42 K to a high-temperature state with integer
valence and conventional electronic mass.
This transition is beyond the scope of the periodic Anderson model, which generally
exhibits gradual changes (crossovers) in its temperature dependence\cite{hewson}. It is, however, reminiscent
of Mott-Hubbard physics, in which a phase transition associated with a strong coulomb
repulsion term in the Hamiltonian can occur.  In this paper we present spectroscopic
studies of the phase transition and the low-temperature state of \ybin\ which show the emergence below the 
transition of a well-defined mode at an energy of about 0.25 eV.  Possible interpretations
of this electronic excitation, and the question of what Hamiltonian is required to capture the
essential physics of \ybin\ are discussed.

The presence of local magnetic moments in metallic systems is 
associated with a variety of interesting
phenomena, including the Kondo effect, heavy-fermion physics and mixed-valence\cite{hewson}.  
In certain cases, an isostructural
first-order transition occurs at which a discontinuous change
in valence and volume is accompanied by an abrupt disappearance
of the local moment\cite{lawrence0,allen2,nowik1,nowik2,sarraop}.  
The transition to such a state provides an opportunity to explore 
fundamental phenomena associated with magnetic moments in metals, moment compensation, 
Kondo singlet formation, and mixed-valence.

At the $\gamma - \alpha$
transition\cite{lawrence0} of Ce a valence change from about 3 to 3.2
occurs in concert with a volume reduction of about 15\% at \Tv $\sim$200 K .
This transition has been discussed in terms of a Kondo-volume-collapse model\cite{allen2}. 
In that model, a reduced lattice constant in the low-temperature phase
is associated with an increase in the hybridization interaction between
local moment and conduction electron states, 
which leads to moment disappearance due to the formation
of a Kondo-singlet state in which
the f-level moments are compensated  
due to a Kondo-like screening by conduction electrons\cite{hewson}.

\ybin\ exhibits an isostructural
transition to a mixed-valent ground state
at which the volume change is almost negligible ($\simeq0.5\%$) 
\cite{kindler,lawrence2,sarrao2,immer,lawrence3}.
At this transition the Yb valence decreases from $\sim$3 to $\sim$2.85,
and the local moment, present in the high-temperature state, vanishes abruptly.
The Kondo scale, $T_K$, which is a measure of the strength of the hybridization
interaction, increase abruptly from $T_K \approx 25 K$ above the transition
to $T_K \approx 400 K$ below the transition for reasons which are not at all clear.

In this letter, we focus on changes in the infrared 
conductivity of \ybin\ associated with the transition to the mixed-valent state.  
The abrupt increase of the Kondo scale below \Tv\ can allow us to identify key features
of the Kondo state, and thus shed light on fundamental phenomena of
Anderson lattice systems. This work is complementary to previous optical work
which addressed the relationship between spectral features and band-structure
calculations\cite{marab,galli,cont1,cont2}.  It is also possible that the very high
purity of present samples\cite{lawrence2,zherl99} allows features to be observed that could not be readily
discerned in earlier work. 

The samples used in these experiments are high-quality single crystals grown from 
an In-Cu flux\cite{sarrao2}.
For these samples a sharp transition occurs at about 42 K in the absence
of strain.  At the transition the volume increases by about 0.5 \% as the sample
is cooled, and the
susceptibility and resistivity drop abruptly by an order of magnitude.
Thermal cycling tends to induce strain in the samples, which can broaden
the transition and move it to higher temperature\cite{sarrao2}.
Infrared and optical measurements are performed using a combination
of Fourier transform and grating spectrometers to cover the range from
50 to 50,000 \wn .  In these measurements we have gone to great efforts to measure in
all ranges before going through the transition to avoid disorder effects
influencing the infrared data significantly.
The conductivity as a function of frequency is obtained from a Kramers-Kronig
transform of the reflectivity data.  
For the purpose of performing this transform, the measured
reflectivity is extended from 50,000 to 200,000 \wn\ as a constant, and above
that it is made to decrease like 1/$\omega^2$.  At low frequency a Hagen-Rubens
termination is attached to the data.  
In the region of the actual data,
the conductivity is insensitive to the details of these terminations.

Figure 1 shows reflectivity and conductivity up to 15,000 \wn . 
These data show the emergence of a prominent
resonance in the mixed-valent (low-T) state
near 2,000 \wn ($\simeq\!1/4 eV$).  The spectral sharpness and the abrupt
appearance of this feature as a function of temperature are striking.

In addition, the data indicate the
persistence of significant temperature dependence 
to relatively high frequency (compared to T or \Tk ) in \ybin .
In the frequency range from about 5,000 to 12,000 \wn\ ,
\sigo\ decreases substantially as T is reduced both above and below \Tv\ .
Recent theoretical\cite{rozenberg,freericks2} and experimental work\cite{fesi,bucher1,bucher3,wachter94,degiorgi99}
has explored the phenomenology and possible origins of such high energy spectral weight shifts 
in strongly correlated systems 
(involving energies vastly larger than $k_{B}T$ and $k_{B}T_K$). 

\begin{figure}[htbp]
\epsfxsize=4.1in
\centerline{\epsffile{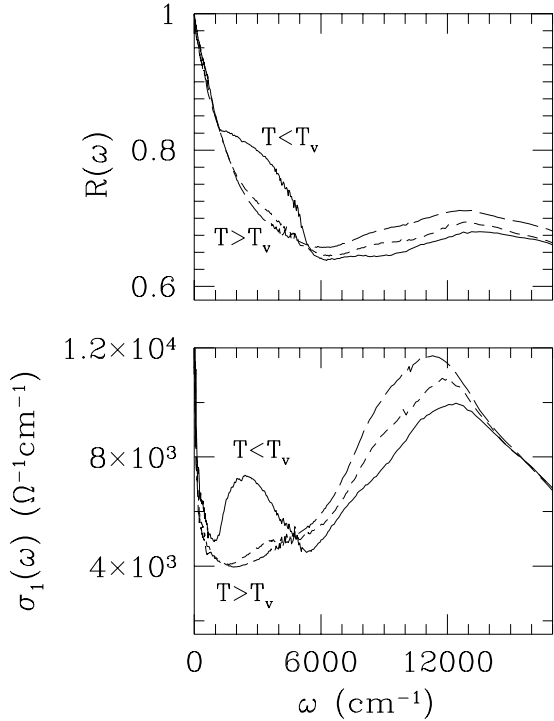}}
\caption{
The reflectivity and the real part of the conductivity at low frequency
are shown for \ybin\ at T= 250 K (long dashes), 55 K (shorter dashes) and 20 K (solid).  
Gradual reduction of spectral weight with cooling occurs in the vicinity of 8,000 \wn\ (1 eV).
A well-defined resonance appears at 2,000 \wn\ (1/4 eV) in the low temperature 
Kondo state.
}
\label{fig1}
\end{figure}

Figure 2 shows spectral weight, which is the indefinite integral of \sigo\ ,
$n(\omega) = \frac{m}{\pi e^2}\int_0^\omega \sigma_1 (\omega^{\prime})\,\mathrm{d}\omega^{\prime}$,
as a function of frequency.
In this figure (and figure 1) we see that there is a net loss of spectral weight
as the temperature is lowered from 250 K to 55 K.  The loss amounts to about 10\% of the strength of 
the broad mode centered around 12,000 \wn , and corresponds to $\sim1.5$ carriers/Yb
atom with the reasonable assumption of a band mass of 4 times the free-electron mass.  
Since spectral weight is ultimately
conserved (if one integrates to high enough frequency\cite{wooten}), 
these data imply that it must be displaced to
still higher frequency (above 16,000 \wn $\simeq2$ eV) as T is reduced from 250 to 55 K.  

On the other hand,
the coalescence of the 20 and 55 K curves at the high frequency end of figure 2  
indicates that spectral weight is conserved within the frequency range from 0 to 15,000 \wn\ as 
the sample goes through the transition.  In particular, the increase in spectral weight associated with
the appearance below \Tv\ of the resonance at $\sim\!2000$ \wn , which
corresponds to about 1.5 carriers/Yb,
is balanced by a general reduction of \sigo\ up to $\sim\!15,000$ \wn .

\begin{figure}[htbp]
\epsfxsize=2.6in
\centerline{\epsffile{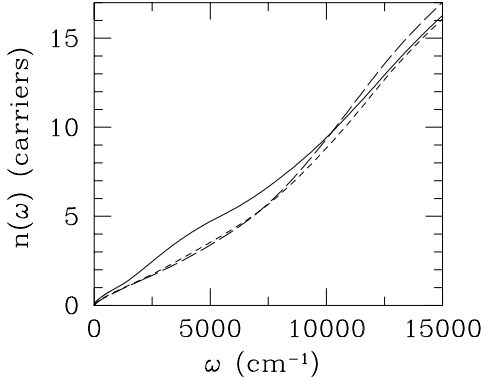}}
\caption{
Spectral weight,
the integral of \sigo\ as a function of frequency from 0 to \w , is shown as a function of
\w\ at T= 250 K (long dashes), 55 K (shorter dashes) and 20 K (solid).  
The units for the vertical axis, expressed in terms of carriers per Yb atom, 
are established by the assumption of a band mass of 4.
}
\label{fig2}
\end{figure}

Both the starting Hamiltonian and the mechanism that drives the transition to the 
mixed-valent state remain areas of active research for \ybin\ .  
With regard to the mechanism, it has been argued that the 
lattice expansion is clearly too small to directly account for the large change in Kondo temperature 
at the transition\cite{sarraot} (from $T_K \simeq$25 K to 400 K). 
The Falikov-Kimball model is capable of producing a quasi Hubbard-like first-order transition, 
and may be relevant to high-temperature properties of \ybin ,
however it ignores hybridization,
which is certainly important in the low-T state\cite{freericks}.
In the mixed-valent state, where the Kondo scale is large,
the dynamics of the periodic Anderson model (PAM) are expected to be relevant.

Within the PAM context, the 1/4 eV excitation can be associated with a quasiparticle
interband transition involving Kondo resonance states 
near \Ef\ \cite{coleman,jarrell}, as illustrated in figure 3, which exist only for $T_K \gg T$.
The abrupt change of \Tk\ at the transition and the abrupt appearance of the resonance 
are consistent with this interpretation.  Note that just above the transition
$T_K / T \simeq 1/2,$ 
whereas just below the transition
$T_K/T \simeq 10.$
The study of the phase transition of \ybin\ ,
at which \Tk\ changes abruptly by an order of magnitude, can thus allow the first clear
identification of this fundamental excitation.

\begin{figure}[htbp]
\epsfxsize=3.6in
\centerline{\epsffile{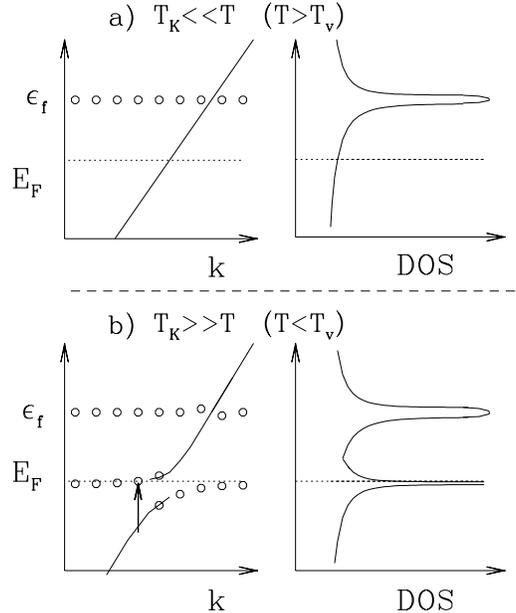}}
\caption{
A generic picture of band dispersion and density-of-states (DOS) for a Kondo-lattice or PAM
system\cite{hewson} is shown for the case of a single f-hole.  
For the E vs k plots, solid lines refer to conduction-band states, and  
circles indicate states with significant f character.
The dotted line is the Fermi level (chemical potential).
a) For \Tk\ $\ll$ T the f level (\epsf\ ), associated with the Yb local moments, 
lies well above the Fermi level (\Ef\ ) and only a rapidly dispersing conduction band crosses \Ef .
b) For \Tk\ $\gg$ T a many-body Kondo hybridization creates slowly dispersing states with
f-like character near \Ef\ (the Abrikosov-Suhl resonance) and an optical transition into such states  
becomes allowed, as shown by the arrow. We propose that the resonance at 0.25 eV in figure 1 may correspond to
this transition.
}
\label{fig3}
\end{figure}

According to model calculations\cite{coleman},
the energy scale for this excitation involving 
dynamically generated quasi-particle states at \Ef\ (the Kondo, or Abrikosov-Suhl, resonance) 
is expected to be $\sim\!\sqrt{T_K B}$. 
Here B is the conduction bandwidth, which appears because of its connection to 
how rapidly the conduction band disperses across \Ef\ ({\it c.f.} figure 3). 
$T_K$ is a measure of the strength of the dynamical hybridization that splits
the bands at \Ef . 

Since $T_K \simeq \tilde{V}^{2} / B$, the infrared determination of this excitation energy
provides a measure of the renormalized hybridization, $\tilde{V}$.
Using the value $\tilde{V}\simeq1/4$ eV from our infrared data, along with
$T_K\simeq400$ K ($\simeq35$ meV), implies a bandwidth of $B\simeq1.8$ eV, which is
reasonable.  One can estimate the hybridization broadening, $\Gamma$,
using its relationship to $\tilde{V}$, to be $\Gamma\simeq0.25$ eV. 
Further, one can use $\Gamma$ in NCA formulae\cite{bickers87} involving
\nft\ along with $L_{III}$ edge measurements of valence\cite{sarrao99}
to infer that the f-level(\epsf ) is about 0.5 eV away from the chemical potential.
These values are all quite reasonable for this mixed-valent system.

Figures 1 and 2 show that the growth of the 0.25 eV mode is associated with a 
reduction of spectral weight at frequencies up to about 12,000 \wn\ (1.5 eV).
This observation--that the growth of the resonance at $\simeq$1/4 eV comes from
a redistribution of spectral weight from essentially the entire bandwidth--may
have implications for questions related to exhaustion and the time scales relevant to 
screening in Kondo lattice systems\cite{jarrell,millis}.
Does it suggest that conduction electrons
further than $k_B T_K$ from the chemical potential are significantly 
involved in screening in the Kondo lattice?
Further work can be expected to address such questions.
It is also intriguing to note that
an excitation of similar frequency is present in \ybb\ \cite{okamura}, 
for which $T_K \simeq 300$ K, and that related features may also be present in 
spectra from mixed-valent Ce compounds\cite{bucher3}.

In conclusion, we have examined the infrared conductivity of \ybin ,
which exhibits an electronic phase-transition (from a magnetic state to
a non-magnetic, mixed-valent state) at $T_v \simeq$ 42 K.  The most striking feature
of the data is the emergence of an electronic excitation at $\sim$0.25 eV in the 
low-temperature state, which we discuss in terms of excitations into the dynamically
generated Kondo (Abrikosov-Suhl) resonance at the Fermi surface.
Like the sudden reduction of resistivity and loss of local moment susceptibility,
the emergence of this feature appears to provide a direct signal of the onset of coherence
in this strongly correlated electron system.

Acknowledgements:  The authors acknowledge valuable conversations with
D. L. Cox, J. W. Allen, J. K. Freericks, S.A. Kivelson, D. H. Lee and A. P. Young,
and technical assistance from T. Lorey, S.L. Hoobler, and P.J. Kostic.
Work at UCSC supported by the NSF through grant\# DMR-97-05442.
and DMR-00-71949.
Work at Los Alamos is performed under the auspice of the U.S. Dept. of
Energy.  NHMFL is supported by the NSF and the state of Florida.  ZF and
JLS also acknowledge partial support from the NSF under grant \# DMR-9501529.
YWR acknowledges support from the ``Marilyn C. Davis
Endowment for Re-entry Women in Science.''

\bibliography{zs-short,ybcu,valence,kondo99}
\end{document}